\documentclass[USenglish,twocolumn]{article}

\usepackage[utf8]{inputenc}%(only for the pdftex engine)
\usepackage[big]{dgruyter}
\usepackage{graphicx}
\usepackage{natbib,twoopt}
\begin{document}

  \articletype{Research Article{\hfill}Open Access}

  \author*[1]{Ulrich Heber}

\author[2]{Andreas Irrgang}

\author[3]{Johannes Schaffenroth}

  \affil[1]{Dr. Remeis-Sternwarte \& ECAP, University of Erlangen-N\"urnberg, E-mail: ulrich.heber@fau.de}

  \affil[2]{Dr. Remeis-Sternwarte \& ECAP, University of Erlangen-N\"urnberg, E-mail: andreas.irrgang@fau.de}
   
   \affil[3]{Dr. Remeis-Sternwarte \& ECAP, University of Erlangen-N\"urnberg, E-mail: johannes.schaffenroth@fau.de}

  \title{\huge Spectral energy distributions and colours of hot subluminous stars}

  \runningtitle{SEDs and colours of sdB stars}

  %\subtitle{...}

  \begin{abstract}
{Photometric surveys at optical, ultraviolet, and infrared wavelengths provide ever growing datasets as major surveys proceed.
Colour-colour diagrams are useful tools to identify classes of stars and to provide large samples. Combining all photometric measurements of a star into a spectral energy distribution will allow quantitative analyses to be carried out. We demonstrate how to construct and exploit spectral energy distributions and colours for sublumious B (sdB) stars. The aim is to identify cool companions to hot subdwarfs and  to determine atmospheric parameters of apparently single sdB stars as well as composite spectrum sdB binaries. We analyse two sdB stars with high-quality photometric data which serve as our benchmarks, the apparently single sdB HD\,205805 and the sdB + K5 binary PG\,0749+658, briefly present preliminary results for the sample of 142 sdB binaries with known orbits, and discuss future prospects from ongoing all-sky optical space- (Gaia) and ground-based (e.g. SkyMapper) as well as NIR surveys. 
}
\end{abstract}
%  \classification[PACS]{}
 % \communicated{...}
 % \dedication{...}

  \journalname{Open Astronomy}
\DOI{DOI}
  \startpage{1}
  \received{..}
  \revised{..}
  \accepted{..}

  \journalyear{2017}
  \journalvolume{1}
%  \journalissue{1}

\maketitle
\section{Introduction}

Optical photometry and spectroscopy provide the observational basis for  astronomy. Ongoing large photometric surveys provide a huge amount of photometric measurements in several optical passbands, which can be used to identify candidate hot subluminous stars, though spectroscopy is needed for proper spectral typing. However, optical photometry is much more than a mere target selection tool. Time-series photometry (light curves) are a crucial ingredient for asteroseismology of pulsating stars and to identify eclipses, reflection effects, and ellipsoidal variations, in compact binaries.
Single-epoch observations, however, provide crucial information as well.  Spectroscopic distances rely on at least one measured apparent magnitude. Ultraviolet and infrared surveys when combined with optical photometry allow us to construct broad spectral energy distributions (SED), which can be used to determine e.g. the effective temperature of a star, to identify an infrared excess hinting at the presence of a cool companion, and to quantify interstellar absorption from UV flux depression. 

Here we shall not address light variation but restrict ourselves to colour-metric properties of hot subdwarf B (sdB) stars.
Several investigations of hot subdwarf stars have made use of single-epoch photometry. 
With the advent of the International Ultraviolet Explorer (IUE) satellite crucial information to study hot stars arose and early attempts to analyse SEDs of sdB stars were carried out by \citet{1984A&A...130..119H}, \citet{1986A&A...155...33H}, and \citet{2001A&A...368..994A}  % Aznar Cuadrado \& Jeffery (2001)
 by combining low resolution UV spectra from IUE with optical photometry.  %Heber et al. 1984 and Heber, 1986
Colour-colour diagrams combining infrared and optical magnitudes are important tools to identify composite objects, such as sdB stars with F/G/K companions
\citep[see e.g.][]{2003AJ....126.1455S,2008ASPC..392...75G}.

%Also spectral energy distributions of such systems constructed from optical in %infrared photometry have been analysed \citep[e.g.][]{}, most recently by  
%\citet{2017arXiv171007327D}. % Deca et al. PG1018 optical and IR

%\citet{2018MNRAS.473..693V} %Vos et al. spectrum synthesis
 
Subdwarf B stars are core helium burning stars of half a solar mass and in the Hertzsprung Russell diagram they form the extreme horizontal branch. Because radial velocity surveys have shown that the fraction of close binaries amongst single-lined (SB1) sdB stars is as high as 50 \%, common envelope evolution plays an important role in the formation of sdB stars. About 30\% of the sdB stars show composite colours, that is they have companions of spectral types F, G, or K. In many cases the companions to SB1 systems have been found to be white dwarfs, but low mass main-sequence stars (spectral type M) and brown dwarfs have been found as well \citep[see][for reviews]{2009ARA&A..47..211H,2016PASP..128h2001H}. Because it is often very difficult to clarify the nature of the companions from optical data alone, broad SEDs 
are the method of choice to constrain the companions' properties and constrain their nature.

We describe a method to construct broad SEDs by combining measurements in various photometric systems from the ultraviolet to the infrared in Sect. \ref{sect:data}. Synthetic photometry for various photometric systems are calculated from grids of appropriate model atmospheres (Sect. \ref{sect:synthesis}). An objective method to derive various parameters from observed SEDs is presented in Sect. \ref{sect:methodology} and preliminary results are discussed in Sect. \ref{sect:results}. We conclude with an 
outlook.

\section{Constructing observational SEDs}\label{sect:data}

The SEDs of the program stars were constructed from photometric measurements ranging from the ultraviolet to the infrared collected from literature. To eliminate the steep slope of the SED we plot the flux density times the wavelength to the power of three (F$_\lambda \lambda^3$) as a function of wavelength throughout this paper.

\subsection{Photometric data}

\begin{figure}
\includegraphics[width=\linewidth]{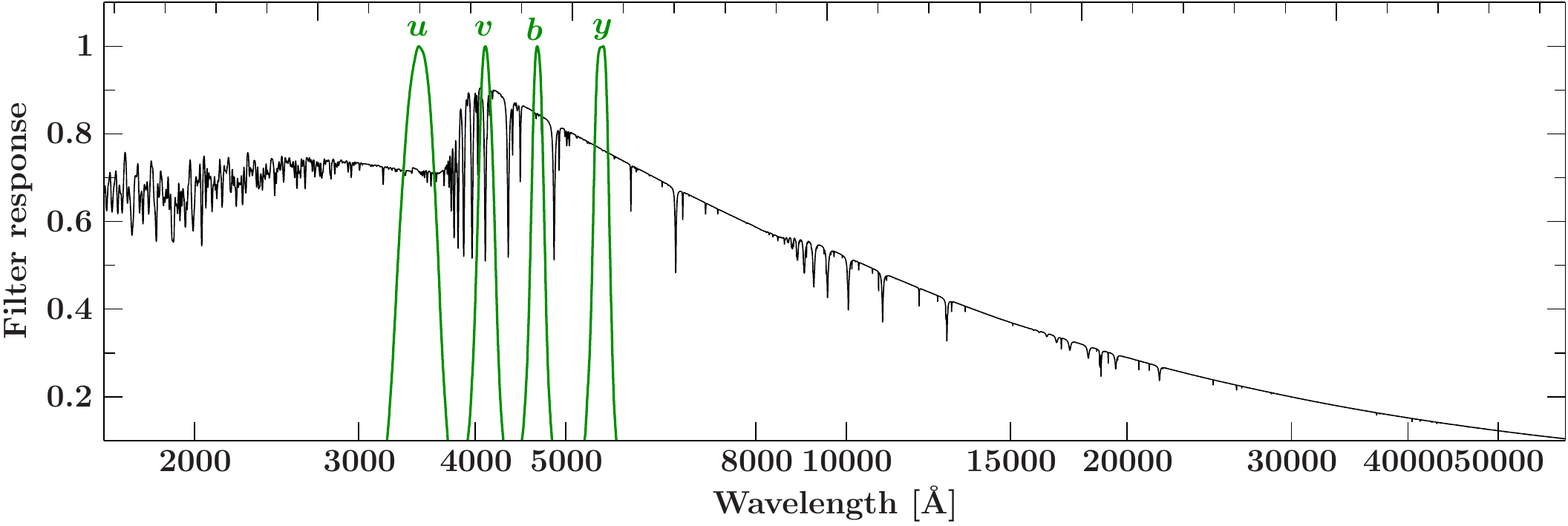}
%\caption{Photometric filter systems: %Str\"omgren}%\label{fig:photometry_stroemgren}
%\end{figure}

%\begin{figure}
\includegraphics[width=\linewidth]{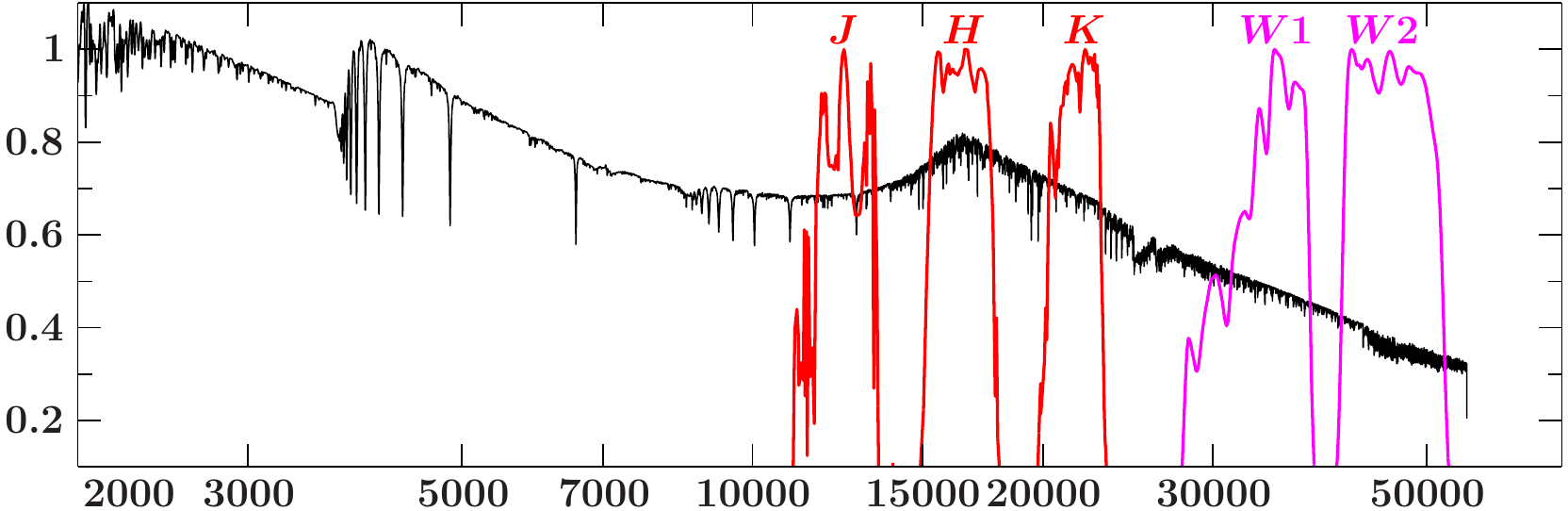}
\caption{Response function for optical (Str\"omgren, top) and infrared (2MASS and WISE, bottom) photometric systems Synthetic SEDs are plotted as F$_\lambda \lambda^3$. In the top panel a single SED is plotted, while for the bottom panel a composite SED is used. Both are scaled for comparison to the transmission functions.}\label{fig:photometry}
\end{figure}

The visual range is covered by SDSS \citep{2015ApJS..219...12A} and APASS \citep{2016yCat.2336....0H} data as well as magnitudes and colours in the Johnson-Cousins, Str\"omgren (see Fig. \ref{fig:photometry}), and Geneva systems, which are collected using Vizier  
%(The General Catalogue of Photometric Data\footnote{\url{http://obswww.unige.ch/gcpd/gcpd.html}} by %\cite{1997A&AS..124..349M}, 
as well as the Subdwarf Database\footnote{\url{http://catserver.ing.iac.es/sddb/}} by \cite{2006BaltA..15...85O}.

Ultraviolet fluxes are important to constrain the atmospheric parameters of a hot subdwarf and were extracted from observations by the International Ultraviolet Explorer (IUE), available in the MAST\footnote{\url{http://archive.stsci.edu/}} archive.

Infrared photometry is of particular importance for binary systems that contain a cool companion because an IR excess is expected.
Available infrared data were taken from ALLWISE \citep{2010AJ....140.1868W,2013yCat.2328....0C}, 2MASS \citep{2006AJ....131.1163S}, and UKIDSS \citep[][see Fig. \ref{fig:photometry}]{2007MNRAS.379.1599L}.  

The photometric datasets are inhomogeneous, both with respect to bandwidth as well as to accuracy. Ultraviolet spectra from IUE cover the wavelength range from 1150\AA\ to 3150\AA\ at a spectral resolution of 6\,\AA. The optical spectral range is covered by several filters both narrow band (e.g. Str\"omgren) and wide-band (e.g. Sloan or Johnson), while the infrared is usually represented by five wide-band filters (J, H, K, W1, and W2).

\subsection{Ultraviolet fluxes from IUE spectra}

\begin{figure}
\begin{center}
\includegraphics[width=0.9\linewidth]{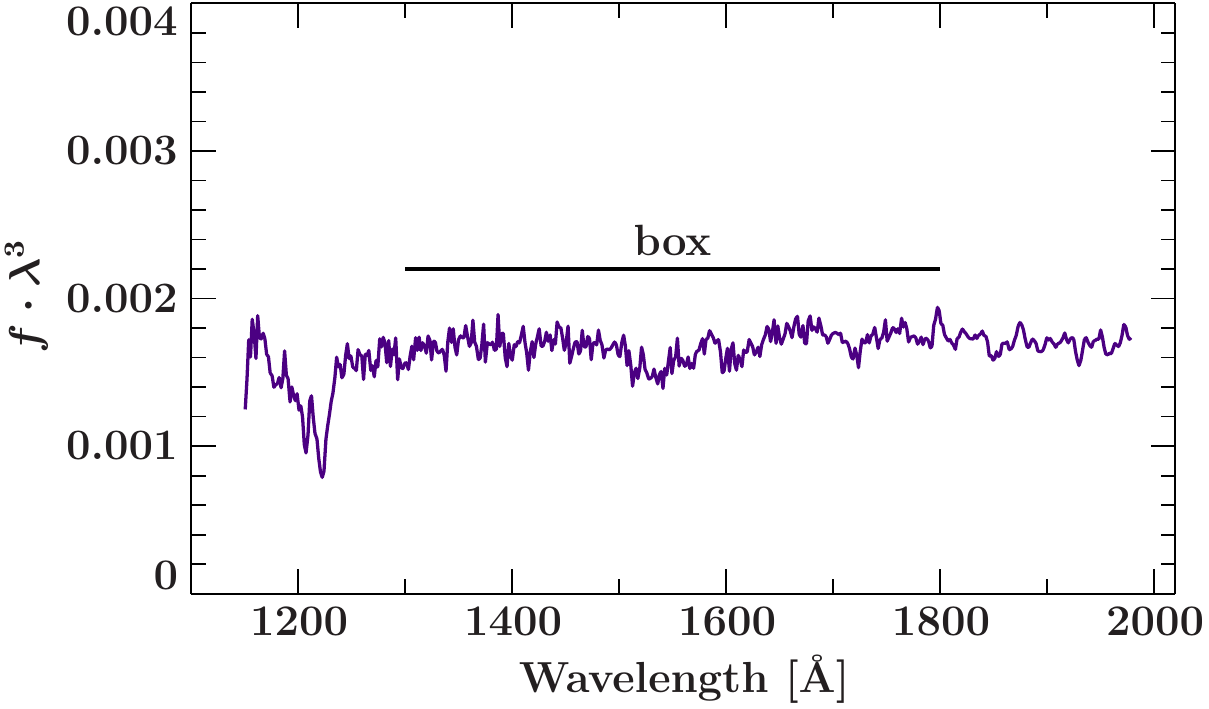}

\includegraphics[width=0.9\linewidth]{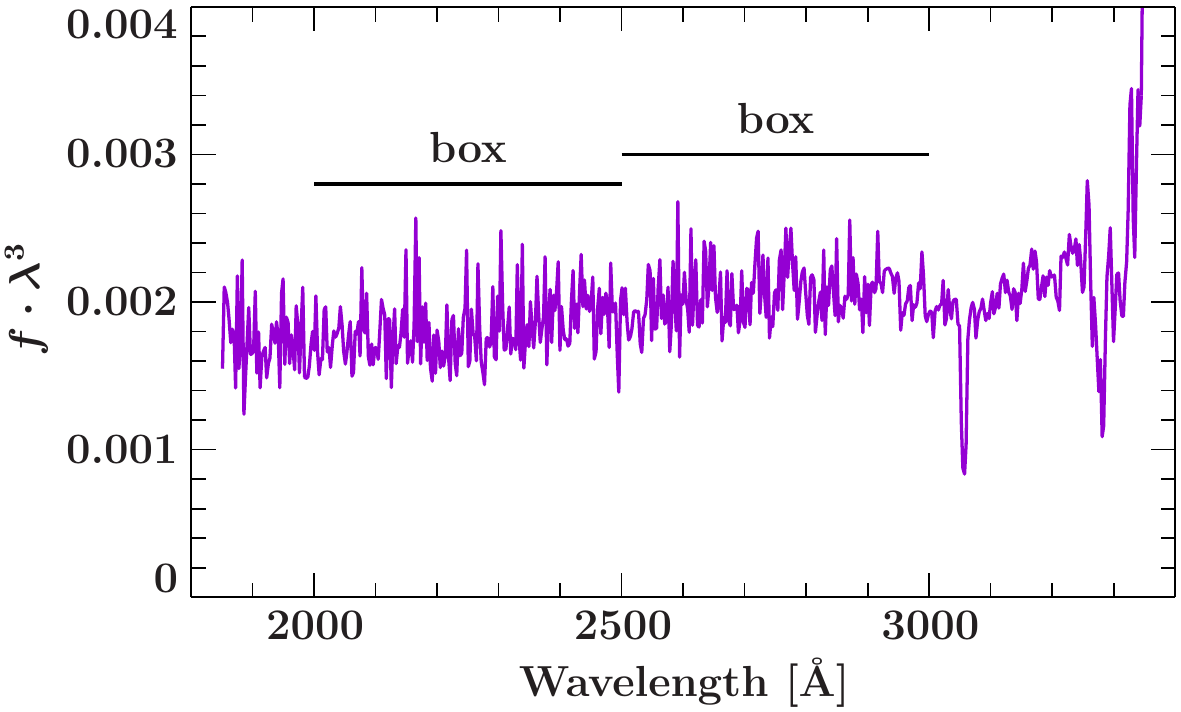}
\end{center}
\caption{Box filters to convert IUE low resolution spectra to UV magnitudes. One box is placed in the wavelength range covered by the short wavelength camera (SWP, upper panel) and two in the regime covered by the long wavelength cameras (LWR/LWP, lower panel).}\label{fig:uv_box}
\end{figure}

The IUE satellite provided UV spectra for two wavelength ranges;
the short (SW, 1150-1975\AA) and long (1910-3150\AA) wavelength range.
%The default cameras were
%%the Long-Wavelength Prime (LWP) and the Short-Wavelength Prime (SWP). The Long-
%Wavelength Redundant (LWR) camera and respectively the Short-Wavelength Redun-
%dant (SWR) would be used if one of the two cameras had failed or would produce errors.
Each spectrograph offered both high and low resolution modes, with spectral resolutions of
0.2 and 6 \AA\ respectively, as well as two entrance apertures each, a
small circular aperture with a 3 arcsec diameter and a large rectangular aperture of 10 by 20 arcsecs. We discarded spectra at high-resolution as well as those taken through a small aperture, because the flux calibration is less accurate than that for
the large-aperture, low-resolution spectra.
Because we have to combine them with broad and intermediate band optical and infrared
photometry we defined a suitable set of filters to derive UV-magnitudes from IUE spectra (see Fig. \ref{fig:uv_box}).
Three box filters, which cover the spectral ranges $1300$--$1800$\,\AA, $2000$--$2500$\,\AA, and $2500$--$3000$\,\AA, are defined to extract magnitudes from the IUE spectra. The box filters were designed in order to avoid the boundaries
of the SW and LW wavelength ranges because of the increasing noise level and the region around
the Lyman-alpha line because of the contribution by
interstellar gas absorption. The mid-UV filter was designed to include the UV absorption bump at $\approx$ 2200\AA\ of interstellar absorption (see Fig. \ref{fig:reddening}), which is important to determine the interstellar reddening parameter E(B-V).  

%For the ultraviolet region, we refrain from using GALEX \citep{2007ApJS..173..682M} observations $%because \citet{2014MNRAS.438.3111C} showed that their absolute calibration is not at the level of %ccuracy required here (see, for instance, the error bars in \cite{2015MNRAS.450.3514K}). 
%
\section{Synthetic SEDs and colours}\label{sect:synthesis}
The magnitude $\mathrm{mag}_x$ of an arbitrary photometric passband $x$ is defined as
\begin{equation}
\mathrm{mag}_x = -2.5\log\left(\frac{\int \limits_0^{\infty}{r_x(\lambda)f(\lambda)\lambda\mathrm{d} \lambda}}{\int \limits_0^{\infty}{r_x(\lambda)f^{\mathrm{ref}}(\lambda)\lambda\mathrm{d} \lambda}}\right) + \mathrm{mag}_x^{\mathrm{ref}}\,\;
\label{eq:definition_magnitude}
\end{equation}

where  $r_x(\lambda)$ is the response function of the filter (see Fig. \ref{fig:photometry} for examples) and $f(\lambda)$ the flux at the photon-counting detector. The flux of a reference star (usually Vega) $f^{\mathrm{ref}}$ is needed to set the zero point of the filter to a predefined magnitude $\mathrm{mag}_x^{\mathrm{ref}}$. 
Note that we assume photon-counting detectors, which explains the additional factor $\lambda$ in the arguments of the integrals \citep[see, e.g.][for details]{1998A&A...333..231B}.

%Based on a model spectrum $F(\lambda)$, which gives the calibrated flux emanating from the surface of %an object, we can exploit Eq.~(\ref{eq:definition_magnitude}) to compute synthetic magnitudes. 
The stellar flux at Earth $f(\lambda)$ can be calculated from the model flux at the stellar surface $F(\lambda)$ and the angular diameter of the star $\Theta$ ($= 2 R_{\star/d}$), which is two times the stellar radius $R_{\star}$ divided by the distance, from which we obtain $f(\lambda) = \Theta^2 F(\lambda)/4$. 
 
%$4\pi d^2 f(\lambda) = 4 \pi R_{\star}^2 F(\lambda)$. 
%To cover deviations from spherical symmetry, we introduce the effective stellar radius, which is %defined by the following equation:%
%\begin{equation}
%4 \pi d^2 f(\lambda) = 4\pi R_{\star\mathrm{,eff}}^2 F(\lambda) \,.
%\label{eq:energy_conservation}
%\end{equation}
%By making use of the angular diameter 

\begin{figure}
\includegraphics[width=\linewidth]{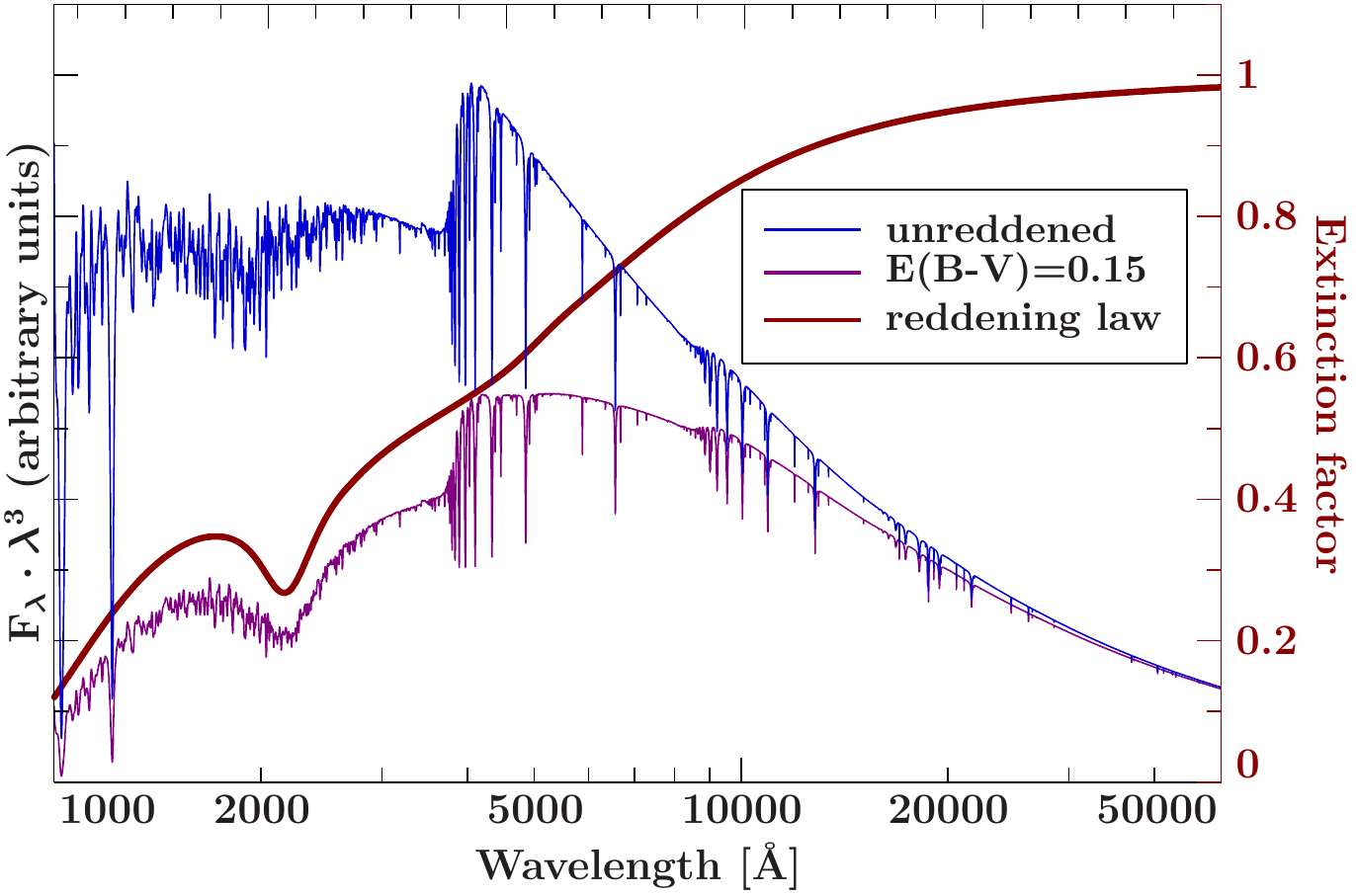}
\caption{The impact of interstellar reddening on SEDs of a B star for moderate interstellar reddening by E(B-V)=0.15 mag. The reddening law is from \citet{1999PASP..111...63F}. }\label{fig:reddening}
\end{figure}

To account for interstellar extinction, the synthetic flux is multiplied with a reddening factor $10^{-0.4 A(\lambda)}$. 
The extinction in magnitude at wavelength $\lambda$, $A(\lambda)$, as a function of the colour excess $E(B-V)$ and the extinction parameter $R_V = A(V) / E(B-V)$ (defaulted to $3.1$) is taken from  \citet[][see Fig. \ref{fig:reddening}]{1999PASP..111...63F}. The final expression to calculate a synthetic magnitude, therefore, reads as%
\begin{equation}
\mathrm{mag}_x = -2.5\log\left(\frac{\Theta^2 \int \limits_0^{\infty}{r_x(\lambda)10^{-0.4 A(\lambda)} F(\lambda) \lambda\mathrm{d} \lambda}}{4 \int \limits_0^{\infty}{r_x(\lambda)f^{\mathrm{ref}}(\lambda)\lambda\mathrm{d} \lambda}}\right) + \mathrm{mag}_x^{\mathrm{ref}}\,.\;
\label{eq:synthetic_magnitude}
\end{equation}

%\begin{itemize}
%\item{\textbf{sdB:}}
%\begin{itemize}
%\item{Model atmospheres: ATLAS9 \& ATLAS12 (Kurucz)}
%\item{Modifications:}
%\begin{itemize}
%\item Level dissolution (Hubeny et al. )
%\item NLTE departure coefficients for hydrogen (20 levels) and He II (x levels) %from DETAIL
%\end{itemize}
%\item{Abundances (scaled ''Naslim''), ''metallicity'' z}
%\item Grid:
%\end{itemize}
%\item{\textbf{Cool star:}}
%\begin{itemize}
%\item{Effective temperature: T$_{\rm eff,2}$}
%\item{Surface gravity: log g$_2$}
%\item{Metalicity: [Fe/H]}
%\end{itemize}
%item{Surface ratio: S}
%\end{itemize}

%sdB  20000-50000  4.6-6.2  

\subsection{Grids of synthetic SEDs and colours}\label{sect:grids}

We aim at modelling the observed SEDs and colours of single sdB stars or SB1 binaries, as well as composite spectrum systems consisting of a hot subdwarf and a late-type main-sequence star.

\subsubsection{SEDs of hot subdwarf stars}

Subluminous B stars are known to show peculiar chemical abundance patterns \citep{2009ARA&A..47..211H,2016PASP..128h2001H}
% Heber reviews
characterized in general by depletions of light metals (C to Ca) and enrichment of heavy metals by very large factors with respect to solar composition. However, star-to-star scatter is large. \citet{2013MNRAS.434.1920N} suggested an average abundance pattern, which we adopted for the model calculations. For the elements not listed in \citet{2013MNRAS.434.1920N}
%Naslim et al. (2013) 
the reference abundance is solar \citep{2009ARAA..47..481A}. %((Asplund et al., 2009).

\begin{figure}
\centering
\includegraphics[width=\linewidth]{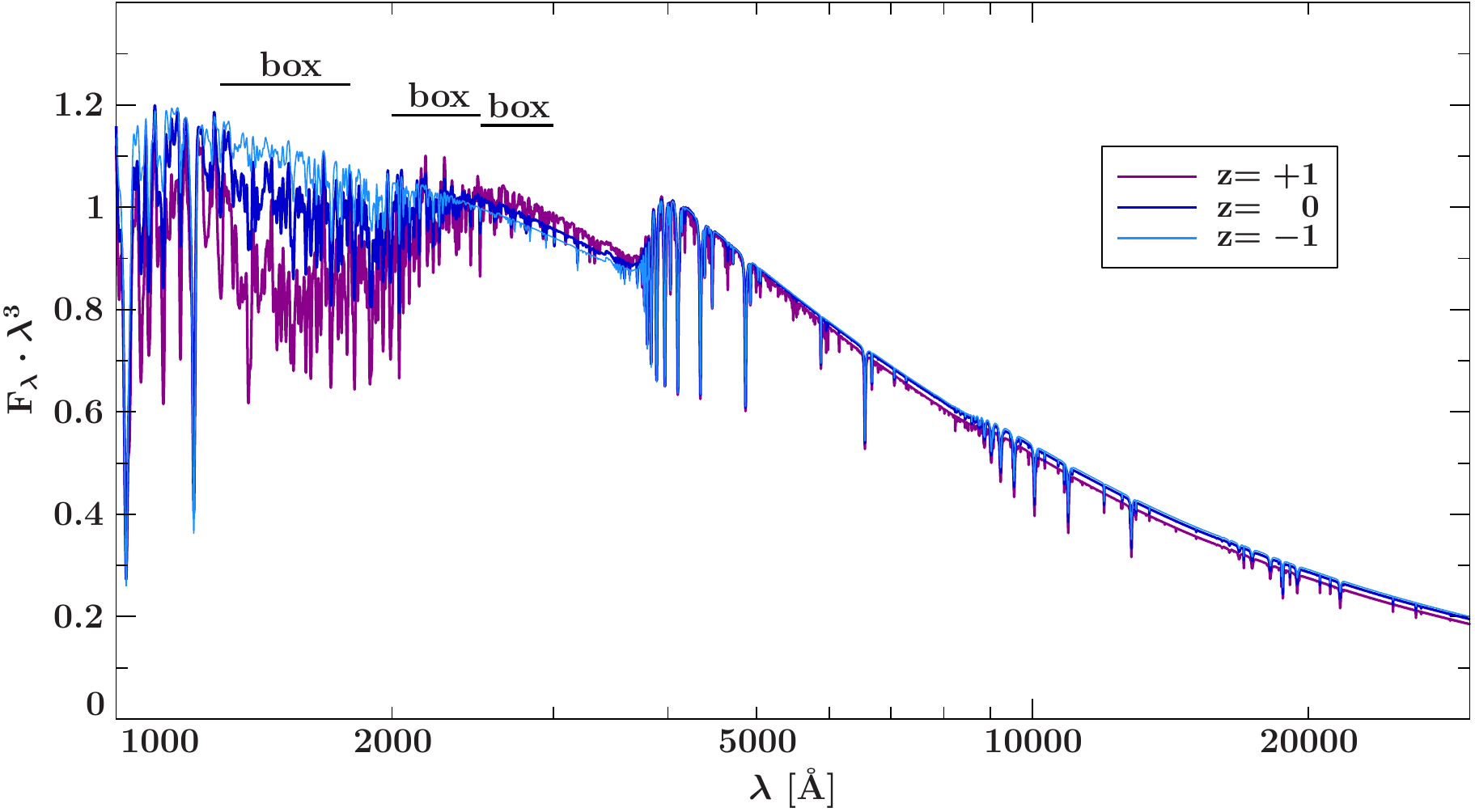} 

\caption{
Model SED for different metallicities \citep[scaled typical sdB pattern from][]{2013MNRAS.434.1920N}.
The UV box filters are marked by horizontal lines. Note that the 
flux depression is strongest in the shortest wavelength box filter.
}\label{fig:metalicity}
\end{figure}

In order to synthesize the SED of a hot subdwarf star a 
grid of model atmospheres was calculated using the ATLAS12
code \citep{1996ASPC..108..160K} % Kurucz 1996
with effective temperatures ranging from 15000 K to 55000 K and surface
gravities from 4.6 to 6.2. The helium abundance was fixed at a low value of one hundredth solar and the logarithmic ''metallicities'' $z$ are scale factors with respect to the abundance pattern of \citet{2013MNRAS.434.1920N}. The synthetic spectra cover the wavelength range form 300 \AA\ to 100\,000 \AA\ (far UV to mid infrared).
The logarithmic metallicity $z$ is allowed to  vary between -1 and +1 (a tenth or ten times the
typical composition of a subdwarf B star). Please
note that iron and nickel are the dominant absorbers and have the greatest influence on
the metallicity z
because they have many absorption lines in the FUV and their absolute abundance
is high.  
%Very heavy elements like zinc or lead are not treated by ATLAS
%because their absolute abundance is so low. 
As demonstrated in Fig. \ref{fig:metalicity}, the metallicity essentially affects the UV spectral range, most significantly the short wavelength UV box filter. Hence it might be possible to derive the metallicity of the sdB if such UV measurements were available.

%An example where the metallicity could be
%determined quite well is discussed in section 4.5.
%The helium abundances was set to a very low value (one hundredth solar) as typical for a sdB star.
% in order to
%avoid artefacts caused by He II absorption edges in the UV for the hotter %models. This
%barely changes the results of the analysis.

Recently, several improvements have been implemented in the ATLAS12 code (Irrgang, in prep.), the most important of which is the treatment of high series members of the hydrogen and ionized helium
line series, following \citet{1994A&A...282..151H}. %Hubeny et al. (1994)
This is of particular importance to model the Balmer jump.

\subsubsection{SEDs of cool stars}

For cool stars a grid of PHOENIX models calculated by \citet{2013A&A...553A...6H} is used % Husser et al. 
 \footnote{\url{http://phoenix.astro.physik.uni-goettingen.de/}}.  The synthetic SEDs cover the wavelength range from 500--55000 \AA. The parameter range is confined to effective temperatures between 2300\,K and 12000\,K, surface gravities between 2 and 5 dex, and
the helium content is set at the solar value.

\subsubsection{Combining SEDs of sdB and cool stars}

In order to combine the spectra of the two components the surface ratio S needs to be determined, which adds another parameter, from which the angular diameter of the companion $\Theta^{\rm c}$ can be derived.

\begin{figure}
\includegraphics[width=\linewidth]{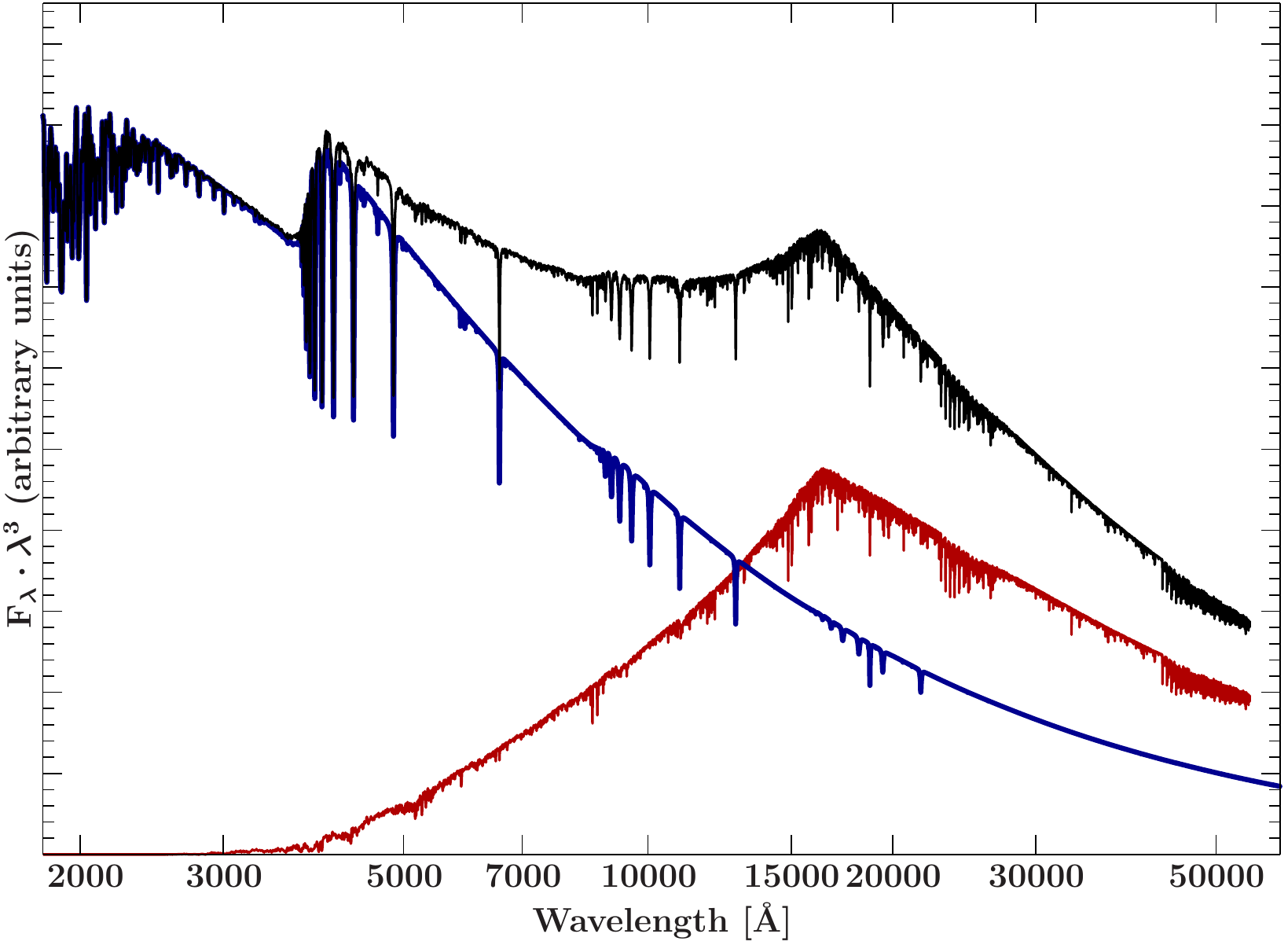}

\caption{Synthetic SED for a composite-spectrum sdB binary. The sdB model (blue) is computed with ATLAS12 for  
T$_{\rm eff}$=30000K. The flux distribution of the cool companion (red) is taken from the PHOENIX grid for a T$_{\rm eff}$=5000K \citep{2013A&A...553A...6H}.
} \label{fig:synthetic_binary_SED}
\end{figure}

\section{Photometric analysis methodology}\label{sect:methodology}

To facilitate objective and efficient photometric analyses, we have developed a grid-based fitting routine. It is based on $\chi^2$ minimization tools provided by the Interactive Spectral Interpretation System \citep{2000ASPC..216..591H} to find the global best-fit in the multi-parameter space.
%, which is normally spanned by the star's atmospheric parameters as well as by %$\Theta$, $E(B-V)$, and --~if necessary due to peculiar reddening~-- $R_V$. The %atmospheric properties of the companion and the surface ratio 
%$A_{\mathrm{eff,c}}/A_{\mathrm{eff,sd}}$ are additional parameters when the %composite SED of a binary star is analyzed.

The six parameters considered to model SEDs and colours of the sdB stars are
%\footnote{In principal other model parameters such as microturbulence, rotation and the helium abundance could be included in the fit, but are insignificant to model sdB SEDs and colours. For helium-rich sdO stars, this helium abundance is important, but the effect needs yet to be evaluated.} 

\begin{itemize} 
\item the angular diameter $\Theta$
\item the effective temperature: T$_{\rm eff}^{\rm sdB}$
\item the surface gravity: log g$^{\rm sdB}$
\item Helium Abundance: $\log$(n(He)/n(all))%
\item ''Metallicity'' $z$ (scaled typical abundance pattern \citep{2013MNRAS.434.1920N})
\item the interstellar reddening parameter E(B-V)
\end{itemize}

Adopting the canonical mass for the sdB star, we also derive the stellar distance.

\subsection{Composite spectra}

In the case of binary stars we may observe a composite spectrum, which increases the parameter space by the parameters describing the companion as well as the surface ratio S of both stars.

\begin{itemize}
\item{Effective temperature: T$_{\rm eff}^c$}
\item{Surface gravity: log g$^c$}
\item{Metallicity: [Fe/H]}
\end{itemize}

However, usually the surface gravity and metallicity of the cool star are unconstrained. Therefore, they were kept fixed to log\,g=4.5 and 1/10 solar metallicity. 

\subsection{Determination of uncertainties}\label{subsection:uncertainties}
Uncertainties are derived from the $\chi^2$ statistics. The parameter under consideration is increased/decreased --~while all remaining  parameters are fitted to account for possible correlations~-- until a certain increment $\Delta \chi^2$ from the minimum $\chi^2$ is reached. The value chosen for $\Delta \chi^2$ determines the confidence level of the resulting interval. For instance, $\Delta \chi^2=1$ yields single-parameter $1\sigma$ uncertainties. 
%A necessary requirement for this interpretation to be valid is that the reduced $\chi^2$ at the best %fit is close to $1$.

The photometric data were compiled from various sources and, thus, are quite inhomogeneous, in particular with respect to the stated uncertainties. 
%Sometimes, the latter are completely missing or different techniques were applied to estimate those that are provided. 
The following strategy was employed to cope with this: (i)~Data flagged in catalogs as uncertain and obvious outliers are omitted. (ii)~Magnitudes and colours without given errors are assigned typical uncertainties of $0.05$ and $0.025$\,mag, respectively. (iii)~To account for systematic shortcomings (e.g.\ in the system response curves, synthetic SEDs, or calibration of the data), a generic error of $0.015$\,mag is added in quadrature to all observed values. (iv)~Eventually, all uncertainties are rescaled by a common factor to ensure a reduced $\chi^2$ of $1$ at the best fit.

\section{Results}\label{sect:results}

We present preliminary results for two sdB stars, the apparently single HD\,205805 and the composite spectrum sdB binary PG\,0749+658. These stars were chosen because extensive, high-quality photometric observations in all relevant wavelength regimes are available. 

\subsection{HD\,205805 -- a benchmark single sdB star}\label{sect:hd205805}

HD\,205805 is one of the brightest sdB stars and, therefore, ample photometric measurements are available for the optical regime including Str\"omgren indices, in particular the H$\beta$ index, which employs a narrow band filter centered at the H$\beta$ line to measure its strength. An ultraviolet spectrum has also been observed by IUE as well as infrared fluxes.
HD\,205805 is one of a handful of sdB stars that have such good photometric data coverage and, therefore provides our benchmark for SED and colour fitting.

High resolution optical spectra taken with the FEROS spectrograph at the ESO 2.2m telescope became also available through the ESO archive. We analysed five FEROS spectra using an updated version of the grid of synthetic hydrogen and helium spectra calculated from metal-line blanketed LTE models described by 
\citet{2000A&A...363..198H} % Heber et al. 2000
The resulting atmospheric parameters are T$_{\rm eff}$=25114$\pm$214\,K, log\,g=4.96$\pm$0.09, and a helium to hydrogen ratio of $\log$(n$_{\rm He}$/$n_{\rm H})$=-1.93$\pm$0.03 by number. 

\begin{figure*}
\includegraphics[width=\linewidth]{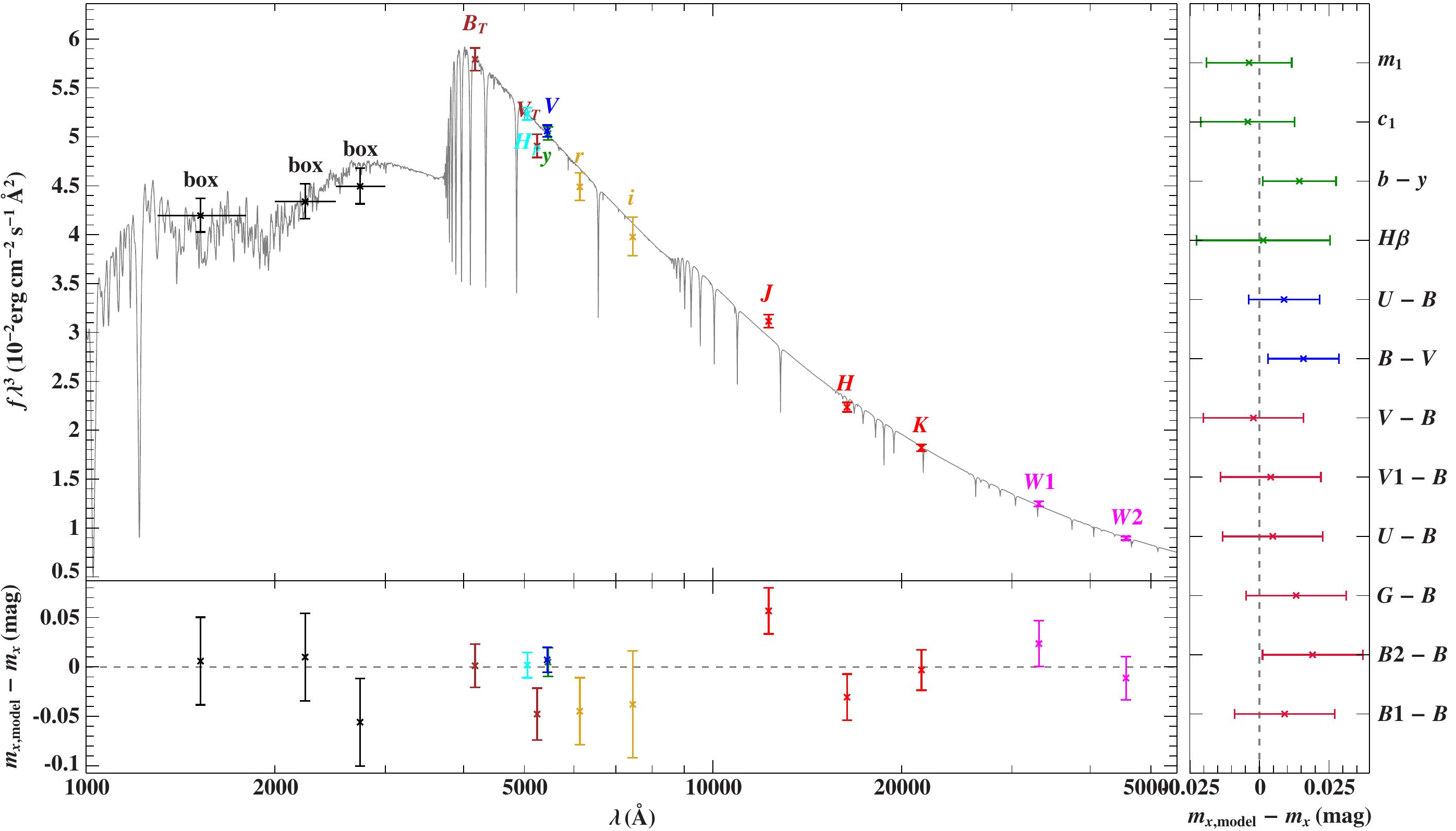}
\caption{HD\,205805: Fit of the SED (top panel) simultaneously with available colours in the Str\"omgren, Johnson, Geneva photometric systems. The magnitudes
B$_T$ and V$_T$ are from the Tycho catalog and H$_P$ is the Hipparcos magnitude.
Residuals are shown in the right hand (colours) and lower (fluxes) panels as magnitudes.}\label{fig:hd205805}
\end{figure*}

\begin{figure}
\includegraphics[width=\linewidth]{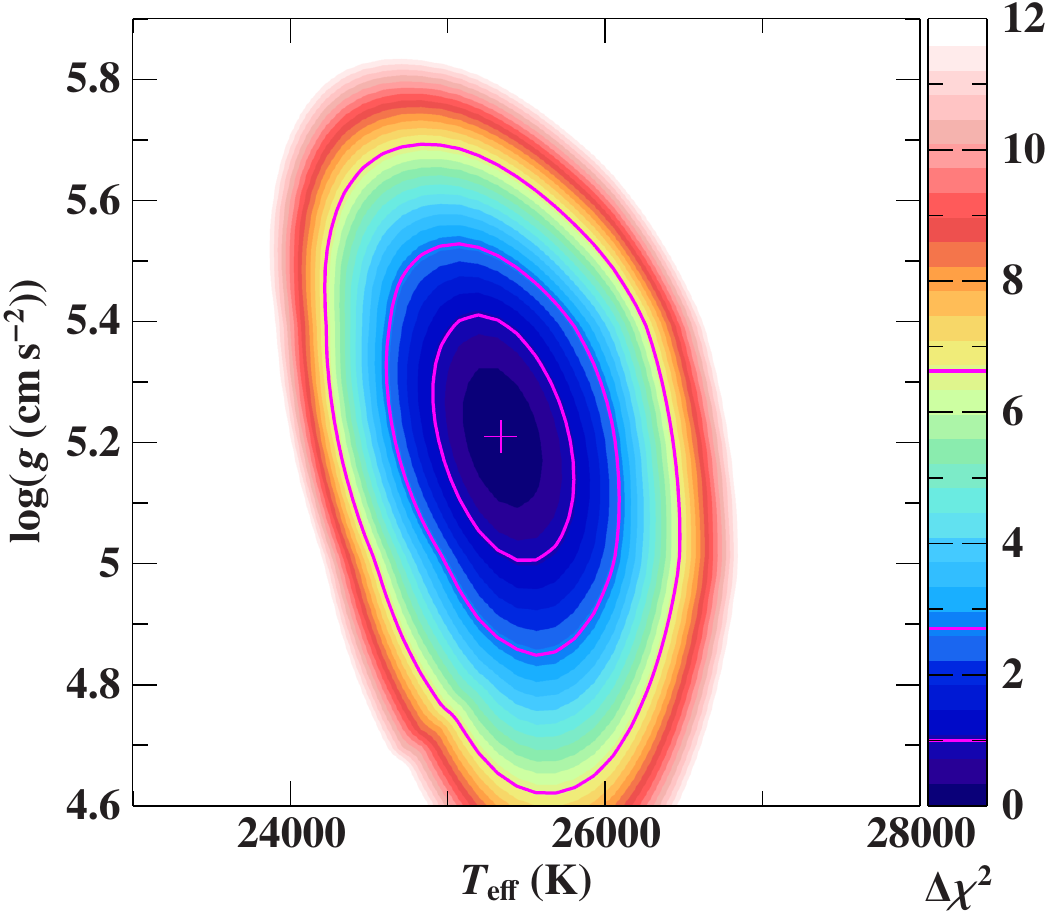}

\caption{HD\,205805: Confidence map for the fit shown in Fig. \ref{fig:hd205805}. The contour lines refer to single parameter uncertainties of 68\%, 90\% and 99\%, respectively.
Single parameter \textbf{1 $\sigma$} uncertainties are $\Delta$T$_{\rm eff}$=400 K, $\Delta$log g = 0.2 dex.
}\label{fig:uncertainties}
\end{figure}

The SED fit and the corresponding confidence map for the error estimation of T$_{\rm eff}$ are shown in Figs. \ref{fig:hd205805} and \ref{fig:uncertainties}, respectively. The resulting parameters (T$_{\rm eff}$=25338$^{+463}_{-423}$\,K, $\log$ g=5.21$\pm 0.21$) are in perfect agreement with those derived from spectroscopy. 
The metal abundance parameter (z=0.09$^{+0.17}_{-0.28}$) points to a normal sdB composition for HD\,205805 consistent with the metal abundances derived by \citet{2013A&A...549A.110G}. The resulting interstellar reddening towards 
HD\,205805 is very low (E(B-V)=0.016 $\pm$ 0.005 mag).

\begin{figure*}
\centering
\includegraphics[width=\linewidth]{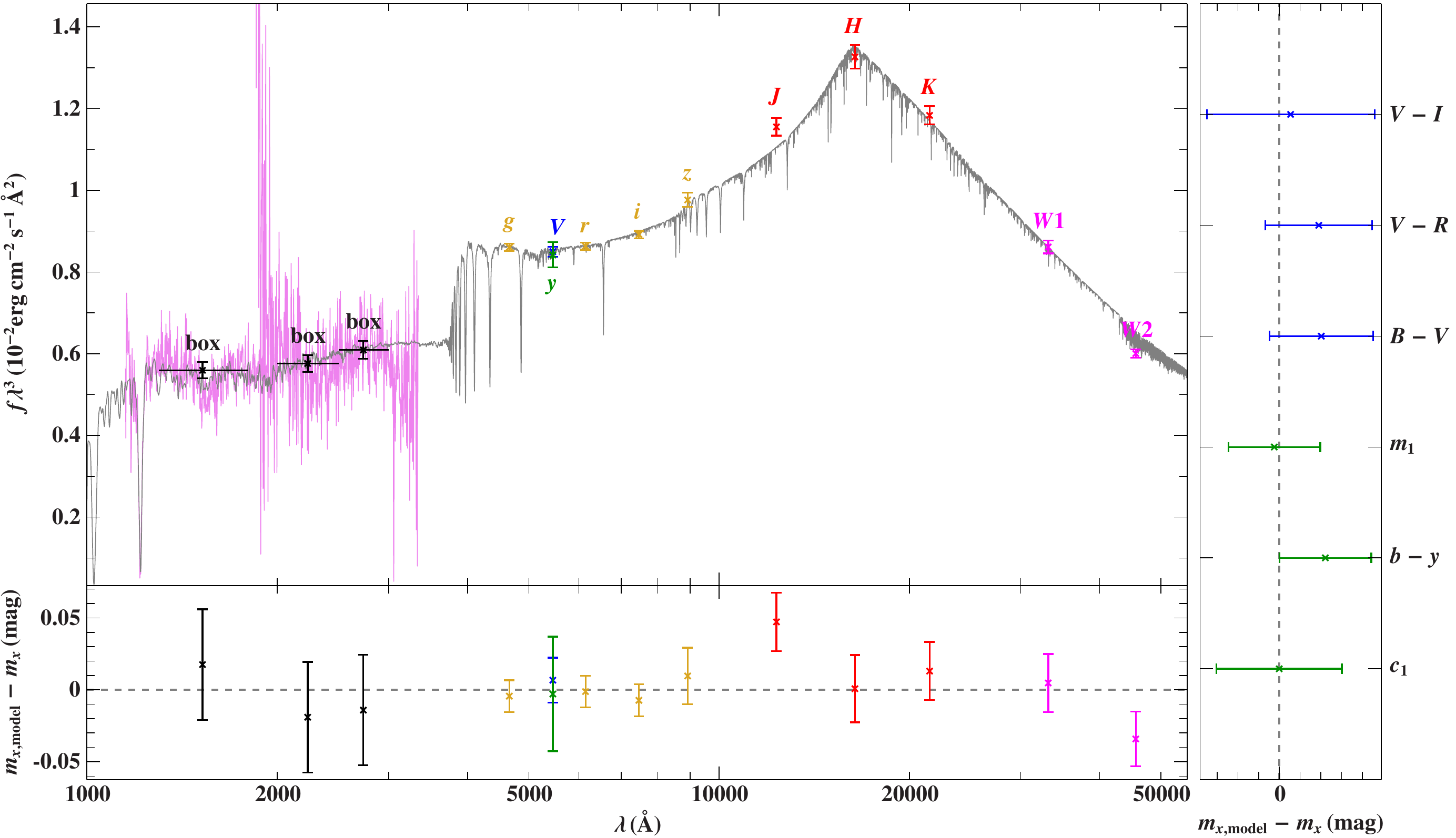}
\caption{
Same as Fig. \ref{fig:hd205805} but for the composite spectrum of PG\,0749+658. The bump in the H band is caused by the onset of H$^-$ absorption.}\label{fig:pg0749}
\end{figure*}

\subsection{The composite spectrum sdB binary PG\,0749+658}

The sdB star PG\,0749+658 was classified as sdB-O by \citet{1986ApJS...61..305G}. % Green et al.
A spectral analysis of the optical spectrum resulted in an effective temperature T$_{\rm eff}$=24600K, surface gravity log\,g=5.54 \citep{1994ApJ...432..351S}. 
% Saffer et al.
 Its composite nature was realised by \citet{1994AJ....107.1565A}  %Allard et al.
 from BVRI photometry and the faint companion was classified as spectral type K5.5, but no significant radial velocity variations were found \citep{2001MNRAS.326.1391M,2002A&A...385..131A}. % Maxted et al. 2001
 
\citet{2001A&A...368..994A} %Aznar-Cuadrado  
determined the effective temperatures for both components from the SED to be T$_{\rm eff}$=25050$\pm$675\,K and T$_{\rm eff}^c$=5600$\pm$300\,K, while \citet{2002A&A...385..131A} analysed the composite spectrum of PG\,0749+658 and derived similar temperatures T$_{\rm eff}$=25400$\pm$500\,K and T$_{\rm eff}^c$=5000$\pm$500\,K. The corresponding gravities were found to be 5.7$\pm$0.11 and 4.58$\pm$0.24 for the sdB and the late-type companion, respectively. 
 \citet{2002A&A...383..938H} % Heber et al. (2002)
derived a considerably lower effective temperature of T$_{\rm eff}$=25050\,K for the sdB component from the optical SED and
attempted to resolve the binary spatially using the Wide Field and Planetary Camera 2 on-board the Hubble Space Telescope, but found it to be unresolved to a limiting angular separation $<$0.2'', which at a distance of 580 pc  translates into a separation $<$ 116 AU. 
\citet{2000ApJ...538L..95O} %Ohl et al. 
determined metal abundances from FUV spectra obtained with FUSE and showed that the sdB is somewhat metal poor in comparison to the typical sdB abundance pattern. 

The fit of the observed SED of PG\,0749+658 is shown in Fig. \ref{fig:pg0749} and the resulting parameters are listed in Table \ref{tab:pg0749}. The resulting temperatures of both stars are lower than those derived from spectroscopy. The resulting gravity of the sdB is consistent with the spectroscopic one derived by \citet{1994ApJ...432..351S} to within error limits, but lower than that of \citet{2002A&A...385..131A}.

\begin{table}
\caption{Results of the analyses of the SED (see Fig. \ref{fig:pg0749}) of PG\,0749+658. The resulting interstellar redening parameter is zero to within error limits (E(B-V)$<$0.01mag.}\label{tab:pg0749}
\begin{tabular}{lllll}
 &T$_{\rm eff} $ [K] & $\log$\,g & $z$ & $\Theta/10^{-11}$ \\
 \hline
sdB &23250$^{+555}_{-391}$ & 5.290$^{+0.19}_{-0.23}$ & -0.16$^{+0.29}_{-0.57}$ & 2.53$\pm 0.07$\\
comp. & \phantom{2}4805$\pm 59 $ & 4.5$^\ast$ & -1$^\ast$ & 10.7$\pm 0.4$\\
%T$_{\rm eff} $& $\log$\,g & $z$ & $\Theta^{\rm c}$ \\
%E(B-V) &  0.00     &    & \\
\end{tabular}

$^\ast$: fixed value
\end{table}
 
\section{Outlook}

HD\,205805 and PG\,0749+658 are amongst the best cases, both in terms of available data quality and wavelength coverage. For most of the other known sdB stars, available datasets are less complete. Hence we can not expect to achieve similar accuracy for the parameters derived from SED fitting, in particular the surface gravity $\log$ g$^{\rm sdB}$ will likely be unconstrained as well as the metal abundance parameter $z$ when no IUE data are available. Because of their large systematic uncertainties, FUV and NUV fluxes from the GALEX mission are not sufficient to replace UV magnitudes from IUE \citep{2015MNRAS.450.3514K}.

Using mock datasets we shall investigate the quality requirements for observed photometric datasets to derive atmospheric parameters to be conclusive.  

\subsection{The sample of sdB binaries with known orbits} 

\begin{figure}
%\centering
%\includegraphics[width=1.0\linewidth]{aa25213-14-fig8.eps}
\includegraphics[width=1.0\linewidth]{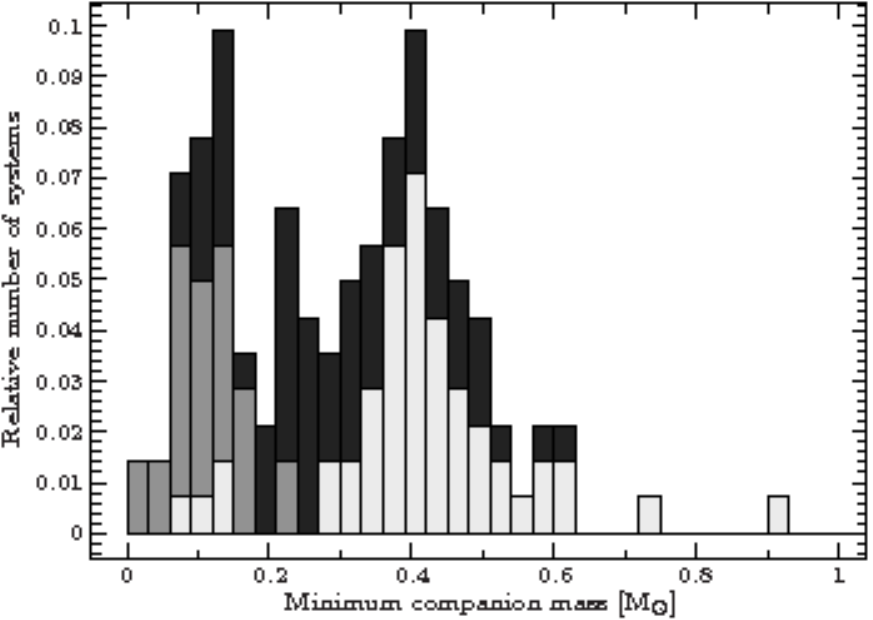}
\caption{Companion mass histogram of the sample of sdB binaries with known orbits \citep{2015A&A...576A..44K}. Systems with white dwarf companions are depicted in light grey, those with M-dwarf companions in grey, and systems for which the nature of the  companion is unclear are marked in black. Adapted from \citet{2015A&A...576A..44K}. }\label{fig:kupfer}
\end{figure}

\citet{2015A&A...576A..44K} % Kupfer et al. 2015
and \citet{2015MNRAS.450.3514K} %Kawka et al. 2015
 compiled a list of close binary sdB stars with known orbits and studied their properties (see Fig. \ref{fig:kupfer}).
 
We restrict ourselves to the single-lined spectroscopic binaries. Because the companions are unseen, they could be white dwarfs, low mass main sequences stars, or substellar objects. From light variations (reflection effect or ellipsoidal variations) and the mass function, the nature of the companions could be inferred only for about half of the sample \citep{2015A&A...576A..44K}.

Hence, we embarked on an analysis of their SEDs in order to better constrain the nature of the companions. We compiled available photometric data from the data archives and constructed the SEDs as described in Sect. \ref{sect:data}. The sample contains 142 stars. Twenty-six are reflection effect systems, hence the companions are normal stars, but were excluded from the study because of their light variability.
\citet{2015A&A...576A..44K} %Kupfer et al. 
suggested that 52 stars host a white dwarf companion. We could model the SEDs of 50 of them by a single synthetic SED, confirming the white dwarf nature of the companion. However, two binaries showed infrared excess and where modelled with a composite SED. The companions are most likely main-sequence stars.
For the stars for which the nature of the companion was unclear, we were able
to reproduce their observed SED with a single synthetic sdB one in fifty cases, but ten binaries require a composite SED, indicating that the companion is likely a late-type main-sequence star. In the case of the single-SED binaries additional modelling is required to possibly clarify the nature of their companions. 
Details will be reported in a subsequent publication. 

\subsection{Gaia, SkyMapper, and other photometric surveys}

Because subdwarf O and B stars are hot the Balmer jump is an important diagnostic tool, which requires measurements of optical UV (e.g. u, u' or U) or NUV magnitudes. Several ongoing surveys will provide such photometric data, in particular SkyMapper, which measures the Str\"omgren u-band, and the Gaia space mission, which will measure spectrophotometry in 30 bands with fine sampling of the Balmer jump. 
All-sky NIR surveys will be important to study composite spectrum sdB binaries and constrain the properties of both components. 

This will put us into an excellent position to constrain the properties of the known ($>$ 5000) hot subdwarfs \citep{2017A&A...602C...2G}, out of which we expect 50\% to be close binaries % Geier et al. (2017)
as well as to enlarge the sample enormously from new discoveries, in particular from Gaia.

\section{Acknowledgements}

Some of the data presented in this paper were obtained from the Mikulski Archive for Space Telescopes (MAST). STScI is operated by the Association of Universities for Research in Astronomy, Inc., under NASA contract NAS5-26555. Support for MAST for non-HST data provided by the NASA Office of Space Science via grant NNX09AF08G and by other grants and contracts. % IUE spectra
This publication makes use of data products from the AAVSO Photometric All Sky Survey (APASS). Funded by the Robert Martin Ayers Sciences Fund and the National Science Foundation. % APASS
This work is based in part on data obtained as part of the UKIRT Infrared Deep Sky Survey. % UKIDSS
This publication makes use of data products from the Wide-field Infrared Survey Explorer, which is a joint project of the University of California, Los Angeles, and the Jet Propulsion Laboratory/California Institute of Technology, funded by the National Aeronautics and Space Administration. % WISE
This publication makes use of data products from the Two Micron All Sky Survey, which is a joint project of the University of Massachusetts and the Infrared Processing and Analysis Center/California Institute of Technology, funded by the National Aeronautics and Space Administration and the National Science Foundation. % 2MASS
This research has made use of the NASA/ IPAC Infrared Science Archive, which is operated by the Jet Propulsion Laboratory, California Institute of Technology, under contract with the National Aeronautics and Space Administration. %

\bibliographystyle{aa}
\bibliography{bib_paper}

%\begin{thebibliography}{} 
 
%\bibitem[Schmidt, 1956]{Schmidt56}Schmidt, M. \textbf{1956}, \textit{Bull. %Astron. Inst. Netherlands}, 13, 15-41.
 
%\end{thebibliography}{} 

\end{document}